# Round-the-world echo of earthquake and solar flare

Anatol Guglielmi[1], Vladimir Parkhomov[2]

[1] Schmidt Institute of Physics of the Earth RAS, Moscow, Russia
[2] Baikal State University, Irkutsk, Russia

**Abstract**

We drew attention to the analogy between the round-the-world seismic echo on the Earth and on the Sun. The phenomenon of echo on the Earth is observed during earthquakes. It was found that the main shock of earthquake excites a circular surface wave, which having rounded the Earth returns to the epicenter and stimulates a repeated earthquake (aftershock) 3 hours after the main shock. Apparently, something similar happens on the Sun during a chromospheric flare. We found signs of a small increase in flare brightness approximately one hour after reaching maximum brightness. The phenomenon is explained by the fact that the converging surface wave, which has returned to the epicenter after the "round-the-world trip", has a definite impact on the flare area.

**Keywords:** earthquake source, main shock, aftershock, chromospheric flare, circular surface wave

## 1. Introduction: Earthquake echo

The circular surface wave is excited after the main shock of the earthquake. Having rounded the Earth, the wave returns to the epicenter about 3 hours after the main shock. This phenomenon is called a round-the-world seismic echo [Guglielmi, 2015].

The impact of echo on the earthquake source can cause a repeated earthquake (aftershock). The point here is as follows. The discontinuity of the rocks at the main shock does not completely relieve the stresses accumulated in the earthquake source. The geological environment in the epicentral zone remains in a non-equilibrium metastable state. Therefore, the round-the-world echo can be a trigger causing a tangible aftershock.

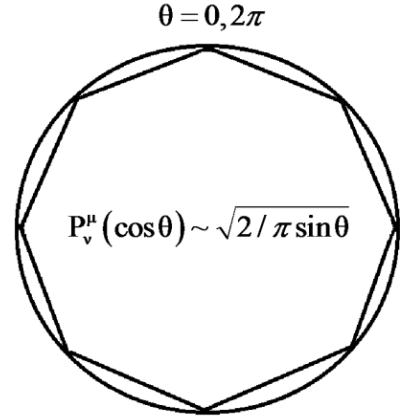

Fig. 1. Schematic picture of round-the-world echo rays created by surface and body waves (smooth and broken lines, respectively).

Figure 1 gives a general idea of a round-the-world echo. We see that the amplitude of the echo is steadily increasing with the approach to the epicenter. Indeed, the Legendre polynomial $P_\nu^\mu(\cos\theta)$ is proportional to the amplitude of oscillations at an angular distance $\theta$ from the epicenter. In the asymptotics $P_\nu^\mu(\cos\theta) \sim \sqrt{2/\pi \sin\theta}$, i.e. the amplitude of the round-the-world echo increases with the approach to the epicenter $\theta = 2\pi$. It is quite clear that the amplitude does not increase to infinity, as indicated by the asymptotic theory in the framework of the spherically symmetric model of the Earth. Due to diffraction, as well as due to spherical and chromatic aberration, the wave amplitudes stay bounded. Nevertheless, it is plausible to assume that the cumulative effect on the earthquake source of converging seismic waves can become an aftershock trigger approximately 3 hours after the main shock.

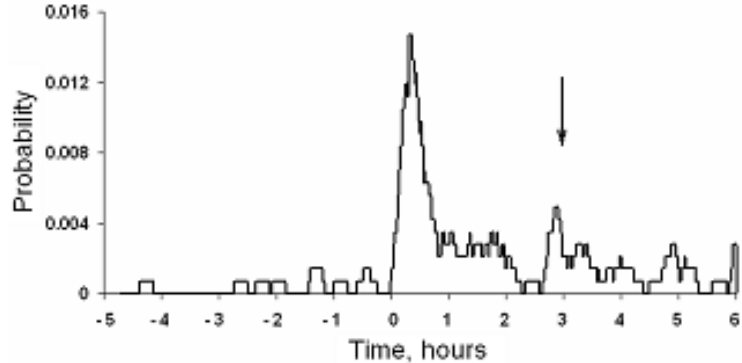

Fig. 2. Dynamics of the aftershocks in the epicentral zones of the strong earthquakes [Guglielmi et al., 2014]. The arrow indicates the expected time delay of round-the-world echo of the surface wave.

A statistical analysis of earthquake catalogs was carried out in [Guglielmi, Zotov, 2012, 2013], and experimental results confirm the hypothesis outlined above (see also [Guglielmi et al., 2014; Zotov et al., 2018]). Figure 2 shows the density of probability of occurrence of the earthquakes, depending on the time. We see the expected aftershock activation about 3 h after the main shock.

Round-the-world seismic echo occurs not only on the Earth, but also on other celestial bodies with spherical symmetry and explosive activity. In particular, the appearance of a round-the-world echo can be expected after a chromospheric flare on the Sun. We attempted to detect a round-the-world echo on the Sun. The result is described in the next section of this paper.

## 2. Echo of a solar flare

So, we want to draw an analogy between the main shock of an earthquake and the chromospheric flare on the Sun. In both cases, there is a rapid release of energy in a relatively small volume, and in both cases circular surface waves are excited, creating a round-the-world echo effect of a one or another intensity. Tentatively the velocity of surface wave on the Sun is in the range of 500 - 1500 km/s. This means that the expected delay of the echo is approximately 35 - 105 minutes. (Recall that the delay of the round-the-world echo on Earth is about 180 minutes.)

There is reason to assume that the high concentration of energy and momentum in a helioseismic wave converging to the epicenter of a flare will cause a transient increase in brightness. It is difficult to expect that the effect will be strong. First, the strong effect of a helioseismic round-the-world echo, most likely, would have already been noticed by someone. More good reason for cautious judgments about the magnitude of the effect is based on data from terrestrial seismology. It is known that the aftershock energy is at least 30 times less than the energy of the main shock of an earthquake (e.g., see the review [Guglielmi, 2015] and the literature mentioned therein). But no matter how weak the effect, its search is interesting from a cosmophysical point of view.

One of the authors (V.P.) studied in detail the effects of chromospheric flares on the plasma shells of the Earth (e.g. see [Parkhomov, 1992; Parkhomov et al., 2006]). During the study, he accumulated a rich archive of basic data on X-ray radiation from the solar flares. This archive we used for the pilot analysis in order to search for a possible echo effect on the Sun. In addition, we used information from https://satdat.ngdc.noaa.gov/sem/goes/data/new_avg/ about GOES observations of X-ray radiation in the wavelength ranges 0.5–4Å and 1–8Å.

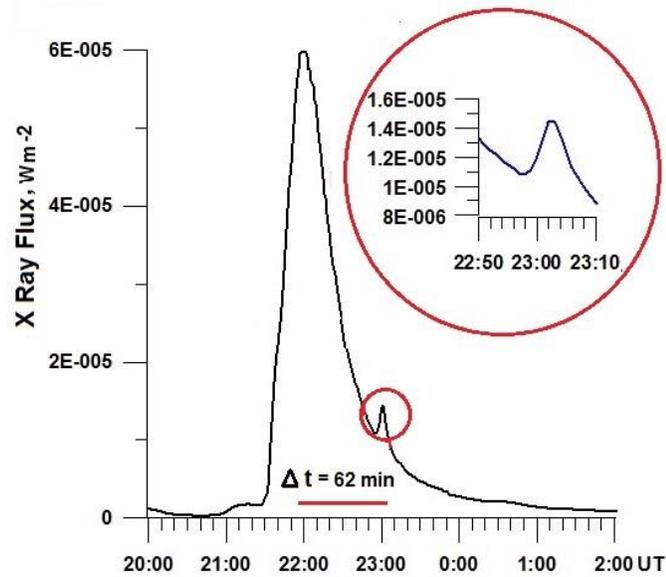

Fig. 3. Chromospheric flare of class X = 2.3 that occurred on 10.09.2005. The X-ray intensity in the wavelength range of 1–8 Å is plotted on the vertical axis. In the upper right corner, a fragment is shown on an enlarged scale, highlighted by a small red circle.

We restrict ourselves in this paper to the description of one event, which illustrate the general idea of our study. Figure 3 shows the dynamics of bremsstrahlung X-rays of the Sun recorded on September 10, 2005 on the GOES satellite in the range of 1–8 Å. We see that 62 minutes after reaching the absolute maximum intensity, a secondary maximum appears, presumably associated with the arrival of the round-the-world echo in the flare region. The delay time corresponds to what we expected. The secondary maximum rises above the current intensity value by about 30%. The height of the secondary maximum is 4 times less than the height of the main maximum.

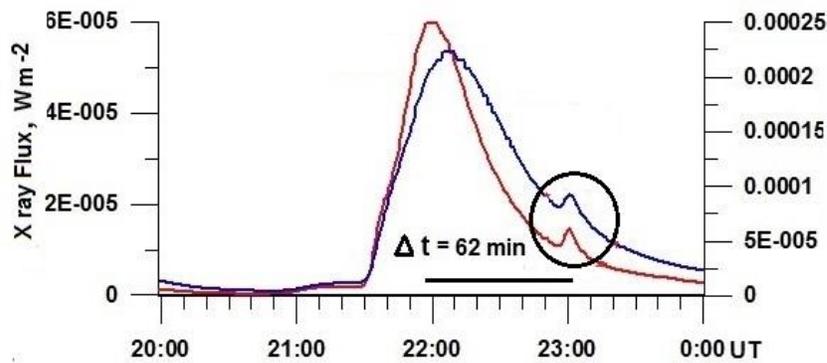

Fig. 4. Same as in Fig. 3, but in two X-ray wavelength ranges: 0.5–4 Å (blue curve, right scale) and 1–8 Å (red curve, left scale).

Figure 4 shows the result of simultaneous measurement of intensity in the two wavelength ranges, 0.5–4 Å and 1–8 Å. We can see that the secondary peaks appear simultaneously in both bands.

It is known that according to data on earthquakes the effect of a round-the-world echo does not always occur on the Earth [Zotov et al., 2018]. Similarly, the echo effect on the Sun is not always observed, but often enough. We noticed signs of an echo in about 10% of the one hundred events viewed by us. The delay time of the echo varies from case to case from 0.5 h to 2 h. The dependence of the delay time on the flare parameters remains to be seen.

After the strongest earthquakes, there is a multiple seismic echo. [Guglielmi, et al., 2014; Zotov et al., 2018]. Occasionally we observed something similar on the Sun. However, the question requires further careful study.

## 3. Discussion and Conclusion

The multiple propagation around the Earth of surface elastic waves has been well known for a long time. This phenomenon relates to the field of linear seismology. In contrast, we focus our attention on the nonlinear phenomenon of the round-the-world echo. This echo stimulates the activity of aftershocks in the earthquake source, that "cools down" after the main shock.

The effect of a round-the-world echo on Earth has been studied in detail statistically and with specific examples. On the Sun, we have so far found only isolated, but rather expressive events. They indicate that the phenomenon of a round-the-world echo of a solar flare deserves the most careful study. We plan to make a statistical analysis of solar flares and present the result in a separate paper.

In conclusion of this paper we will focus on the possible manifestation of an antipodal solar flare effect. The idea is that the surface of the Sun acts as a collecting lens. It concentrates the rays at two antipodal points — at the flare point, and at the diametrically opposite point.

Let's go back to Figure 1. We see that in antipodal point ($\theta = \pi$) the extremal concentration of energy and momentum of the surface wave is expected. Now imagine that the flare occurred on the Sun's limb. Should we not expect that a bright enough spot will flare up shortly afterwards at a diametrically opposite point of the limb? On the Earth an effect of this kind appears in earthquakes, but rather weakly. The antipodal effect is much more pronounced on the Moon and Mercury (see the review [Guglielmi, 2015] and the literature mentioned therein). We plan to search for an antipodal effect on the Sun using the data on chromospheric flares on its limb.

*Acknowledgments*. We express our deep gratitude to O.D. Zotov, who owns the honor of detecting the cumulative effect of seismic waves on the earthquake source. He drew our attention to the circular surface wave arising on the Sun after a chromospheric flare. The authors thank for GOES X rays data which are freely available from the https://satdat.ngdc.noaa.gov/sem/goes/data/new_avg/. The work was supported by the project of the RFBR 18-05-00096, Program No. 12 of the RAS Presidium, as well as the state assignment program of the IPhE RAS. The work of Parkhomov V.A.

was partially supported by the Russian Foundation for Basic Research under the projects No. 18-55-52006 МНТа.

## References

*Guglielmi A.V.* Foreshocks and aftershocks of strong earthquakes in the light of catastrophe theory // Physics–Uspekhi. 2015. V. 58 (4). P. 384–397.

*Guglielmi A, Zotov O.* Impact of the Earth's oscillations on the earthquakes // arXiv:1207.0365v1 [physics.geo-ph]

*Guglielmi A.V., Zotov O.D.* On the near-hourly hidden periodicity of earthquakes // Izv. Phys. Solid Earth. 2013. V..49. No. 1. P. 1–8.

*Guglielmi A.V., Zotov O.D., Zavyalov A.D.* The behavior of aftershocks following the Sumatra–Andaman earthquake // Fizika Zemli. 2014. No. 1. P. 66–74.

*Parkhomov V.A.* Geomagnetic pulsations related with gamma radiation of solar flares // Geomagnetism and aeronomy. 1992. V. 32. No. 1. P. 129–134.

*Parkhomov V.A., Moldavanov A.V., Tsegmed B.* On two different geomagnetic manifestation of solar flare November 4 // Journ. Atmos. Terr. Phys. 2006, 68(12). pp. 1370-1382.

*Zotov O.D., Zavyalov A.D., Guglielmi A.V., Lavrov I.P.* On a possible effect of seismic surface waves traveling around the globe on the dynamic of repeated shocks after large earthquakes // Fizika Zemli. 2018. No. 1. P. 187–201.

**Authors Information**

**GUGLIELMI Anatol Vladimirovich** – Prof., Dr. Sci. (Phys.-Math.), Chief Researcher, Institute of Physics of the Earth, RAS, 123995, 10 B. Gruzinskaya 123995, Moscow, Russia. Ph.: +7 (495) 582-99-71. E-mail: guglielmi@mail.ru

**PARKHOMOV Vladimir Aleksandrovich** – Prof., Dr. Sci. (Phys.-Math.), Professor department of informatics and mathematics Baikal State University, 664003 Irkutsk, Lenin street 11, Russia. Ph.:+7(3952)24-56-93. E-mail: pekines_41@mail.ru